\documentclass[10pt,conference]{IEEEtran}
\IEEEoverridecommandlockouts
\usepackage{cite}
\usepackage{amsmath,amssymb,amsfonts}
\usepackage{algorithmic}
\usepackage{graphicx}
\usepackage{textcomp}
\usepackage{xcolor}
\def\BibTeX{{\rm B\kern-.05em{\sc i\kern-.025em b}\kern-.08em
  T\kern-.1667em\lower.7ex\hbox{E}\kern-.125emX}}

\begin{document}

\title{A Cognitive and Machine Learning-Based Software Development Paradigm Supported by Context 
}

\author{\IEEEauthorblockN{Glaucia Melo, Paulo Alencar, Donald Cowan}
\IEEEauthorblockA{\textit{David R. Cheriton School
of Computer Science} \\
\textit{University of Waterloo}\\
Waterloo, Canada \\
\{gmelo, palencar, dcowan\}@uwaterloo.ca}
}

\maketitle

\begin{abstract}

Advances in the use of cognitive and machine learning (ML) enabled systems fuel the quest for novel approaches and tools to support software developers in executing their tasks. First, as software development is a complex and dynamic activity, these tasks are highly dependent on the characteristics of the software project and its context, and developers need comprehensive support in terms of information and guidance based on the task context. Second, there is a lack of methods based on conversational-guided agents that consider cognitive aspects such as paying attention and remembering. Third, there is also a lack of techniques that make use of historical implicit or tacit data to infer new knowledge about the project tasks such as related tasks, task experts, relevant information needed for task completion and warnings, and navigation aspects of the process such as what tasks to perform next and optimal task sequencing. Based on these challenges, this paper introduces a novel paradigm for human-machine software support based on context, cognitive assistance, and machine learning, and briefly describes ongoing research activities to realize this paradigm. The research takes advantage of the synergy among emergent methods provided in context-aware software processes, cognitive computing such as chatbots, and machine learning such as recommendation systems. These novel paradigms have the potential to transform the way software development currently occurs by allowing developers to receive valuable information and guidance in real-time while they are participating in projects.

\end{abstract}


\section{Introduction}

Software development (SD) is a complex, dynamic process. Many changes are expected to occur during a software development process despite what was planned and what is expected \cite{MEYER2017, Murphy_Beyond2019,Ciccio2015, gronau2005kmdl}. Often, the approach that each software developer takes when performing a task is highly dependent on the characteristics of the software project and its context \cite{Murphy_Beyond2019, murphy_theneed2018}. This context can include technologies (e.g., Eclipse, Jenkins, Java), methodologies (e.g., Scrum), processes (e.g., code, test, deploy), background documentation (e.g., Stack Overflow, API Tutorials), environment (e.g., test, production) and even personal preferences (e.g., Eclipse IDE or VSCode for coding). Given the huge number of possible variations in context, it is cumbersome for developers to remember everything they must do, and how context can influence the way they perform their tasks. For example, in project x, developed in Java with Scrum, two test rounds need to be successful before an automated process deploys project x into production. In project y, developed in Python, developers test the code locally and deploy the code manually by copying some files to a server. Questions developers have include: How long does a sprint take in Project x? Which team/person is responsible for testing each round of project x? Was the test in project x successful? What is the server address to be used for deployment in project y? To which folder should I copy the files? Which files should I copy? and Do I need to keep a version of the files on the server?

Besides the amount of contextual information that remains unmanaged, developers have a high cognitive load remembering the steps for various task workflows. Current efforts to solve these problems are presented through a variety of approaches. An extensive body of research addresses context in software development and includes discussions on many topics including: using context to improve developer productivity \cite{Kersten_Murphy_2006}, providing information to developers \cite{Gasparic_Murphy_Ricci_2017,Holmes_Murphy_2005}, impacting a developer's work through context switching \cite{MEYER2017}. All approaches recognize the rich context inherent in software development and that this context is not always explicit. It has even been argued that context is fundamental to software engineering practices \cite{Murphy_Beyond2019}. 

Another paper has focused on communicating the perceived context with a tool such as a conversational agent \cite{devy}. It is essential to study means to have developers focusing on the creative part of the development, other than investing time in tasks that could be automatically delivered to them through smart systems that understand and use the developer’s current context. There is also work that intends to support the cognitive load of knowledge workers \cite{lima2015, biegel2015}. In all, these papers propose several approaches to support software developers in their daily work activities. 

Thus, SD requires comprehensive support in terms of information and guidance based on context during task execution \cite{devy, Murphy_Beyond2019, MEYER2017}. As a process dependent on knowledge workers \cite{kidd1994marks}, SD lacks supportive methods based on conversational-guided agents that account foesr cognitive tasks such as paying attention, remembering, and maintaining mental maps of the software processes. Current practices of developing software also lack techniques that make use of historical implicit or tacit data to infer new knowledge about the project tasks and navigation aspects of the process. While similar tools and solutions provide comparable assistance \cite{Kersten_Murphy_2006, Kersten2005, Cubranic2004Learning} based on software information, our proposal considers software process information and has a machine learning component. Therefore, we argue that novel approaches should take advantage of the synergy among emerging methods in context-aware software processes, cognitive assistance such as chatbots and recommendation systems based on machine learning (ML). 

Given these challenges, we provoke a discussion by asking:\textbf{ How can software development be advanced by introducing a new paradigm to realize human-machine software development cooperation based on context, cognitive assistance and machine learning? }

SD has already been supported with automated tools \cite{Cub_Murphy_2003, Ponzanelli_Bavota_Penta_Oliveto_Lanza_2014}, and with automatically generated code, commits, and built chatbots \cite{cai2019answerbot, devy}. In the future, developer’s knowledge and ubiquitous context will be integrated into the development environment, complementing the current state-of-the-art with very effective, timely and supportive relevant recommendations for developers. 

This proposed discussion intends to stimulate thinking about the creation of tools and procedures that can advance software development as it relates to software developers and, on a larger scope, companies. There is a direct connection to work being done by large software companies working at the forefront of research and practice involving novel (semi-)automated methods to support the development of software applications while improving software developers’ efficiency and productivity. Shaping software development is critical as software systems have become the backbone of much of today’s technology and society’s functions. Working remotely has become a new reality. Consequently, approaches and tools that help to facilitate software development have become even more essential. Working with software development and its intrinsic implicit context is essential; therefore, we argue there is still the need to improve the machine-developer interaction, instead of purely automating software processes. We discuss the future of this proposed paradigm more in Section \ref{future}. 
 
\section{Motivating Example}

It is known that deployment is a challenging task during software development \cite{carzaniga1998characterization}, and so, we present an example of how changes in the steps that are executed during deployment vary in different contexts. Gabi is a software developer who has been programming in Java for nine years. She has been recently working on project X, a new project for the company. When there is a new project, Gabi needs to create a minimal viable product (MVP) to show her clients. She deploys the software locally, using a container tool such as Docker and manually uploads the project to a web server. She also reboots the server manually after each deployment, so changes are effective immediately. This way is faster, and she does not have to configure a job or a server to generate a deployment automatically, which would cause the clients to wait much longer for the MVP. In contrast, there is project Y, a mature and huge project in the company. When a version of the software in Project Y must go to production, all Gabi does is to commit the code from the local to the shared code repository. Then, the continuous integration pipeline in a Jenkins server will take care of the other steps, which are checking out the code, building the project, uploading the package on the server, and rebooting the server. In theory, the steps are the same, but because the projects and environments are different, Gabi’s work is different. This means that in the second case, the solution should be expecting project information or project phase information such as MVP or production phase, which defines how the form of the deployment. Gabi, who has been working on project Y for years may forget she needs to deploy Project X manually or that the server must be manually rebooted. Having an intelligent cognitive assistant that could support Gabi in many ways by, for example, giving her the necessary information such as the server address, her login information, the folder to which the files can be copied, the link to the deployed system, and the steps to deploy project Y. 

\begin{figure}[h]
 \centering
 \includegraphics[width=0.3\textwidth]{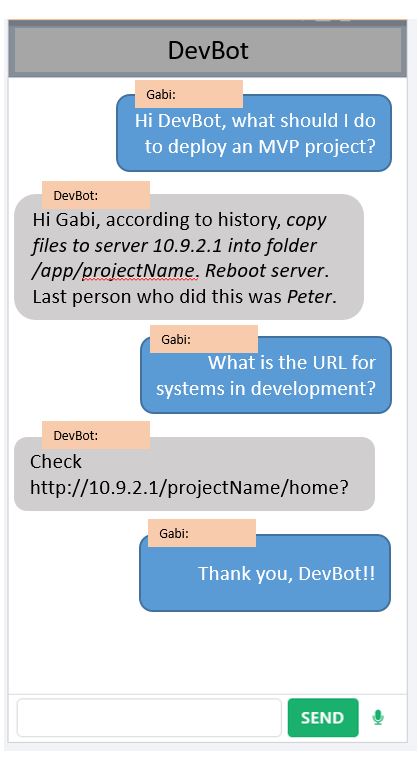}
 \caption{Illustrative chatbot interaction.}
 \label{fig:devbot}
\end{figure}

Imagine Gabi is interacting with a chatbot and does not remember exactly how to deploy a new project, since she is not always working on new projects. In Figure \ref{fig:devbot}, we present a conversation between Gabi and the chatbot during deployment. To reach the desired level of support, the chatbot should know her context, which includes the server address and folder of the development environment, and, based on the history of tasks previously executed in that context, the chatbot should be able to learn and recommend the task sequence that Gabi needs to follow and provide the information required to execute the tasks. 

\section{The Expected Future} \label{future}

\begin{figure}[h]
 \centering
 \includegraphics[width=0.4\textwidth]{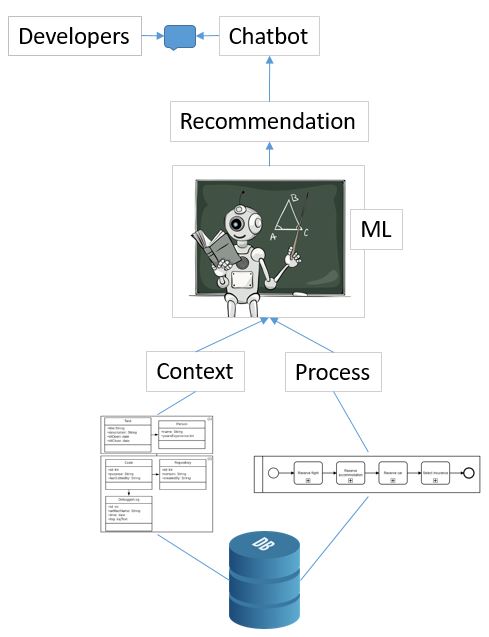}
 \caption{Prototype of conversational channel connecting developer, context and process.}
 \label{fig:architecture}
\end{figure}

In this paper, we envision a new paradigm for chatbot use that knows the software development context and relies on machine learning techniques to support developers when executing their tasks. The purpose of this research is, therefore, to capture the tacit or implicit context and feed it back in a useful way such as making recommendations to developers. Processes based on machine learning and communicated through a chatbot should lower the cognitive load of developers, provide context-aware and real-time support for task execution, and guide developers through the development steps such as deployment. Figure \ref{fig:architecture} illustrates the envisioned architecture of the solution. 

This novel paradigm can potentially transform the way software development is currently undertaken by allowing developers to receive valuable information and guidance in real-time while they are developing their projects. In contrast with the way developers interact with existing IDEs, the proposed paradigm pro-actively provides developers with information and guidance they need, where, when and how they need it. As a consequence, developers will be supported in their interactions, productivity, and decision making.

We anticipate this approach to be deployed in many different software development scenarios, with tools and procedures that explore the reuse of information on software development, providing support and automation to recurrent tasks. Hence, developers can focus on the creative aspects, use of programming languages, data structures, other product-related concerns, and user-focused solutions. The proposed 
paradigm is one step further towards bringing more knowledge to developers, both experienced and novice, during specific software development activities. This paradigm builds upon our previous work on reusing relevant information to support developers in executing their tasks \cite{Glaucia2019} and the relevance of context in software development projects \cite{melo2019context}. 

\section{Making this Future Possible}

To realize a human-machine software collaboration paradigm based on context, cognitive support, and machine learning, requires performing the following research activities.

\textbf{Research on Software Development Context.} We believe it is essential to understand and capture context for software tasks so that adaptive context relevant to software development could be leveraged. Being able to handle this context would allow developers to focus on creative tasks rather than on how to execute a specific procedural task or wonder what should be done next to uncover specific information. The volume of information to be handled, as well as the speed at which such information will arrive, is often massive and rapid. Moreover, the variability of context formats is also an aspect to be considered. Examples of relevant contextual information are the next artifact to be edited or read, an API tutorial, a code snippet, or knowledge from another developer. We strongly believe that the nature of capturing the different contexts and presenting the tasks related to that context is already a significant contribution to the field.

\vspace{0.2cm}
\noindent\fbox{
  \parbox{0.45\textwidth}{
  \textit{Relevant questions include: How can we deepen the use of context to guide developers in real-time? Which context should be leveraged to guide developers in their tasks? How can we better capture and reuse context information in software development? How can massive contextual information be stored so that it can be accessed and used to generate knowledge for software developers?}
  }
}
\vspace{0.2cm}

\textbf{Environment-Developer Interaction.} This approach is within the scope of providing means for developers to interact with a system that is supposed to support them during development. This support is through a conversation where a chatbot supported by a context model should be aware of the developer’s set of tasks. This means the chatbot should know both the workflow (process execution and project characteristics) and the contextual information of the project, as well as be prepared to capture information from developers and other resources providing answers. Bradley and colleagues \cite{devy} have considered a context-model for supporting conversational developer assistants that uses the context elements needed to support workflow involving a distributed version control system. In contrast, the focus of our approach is on guiding developers in the steps they must perform throughout the development project, considering different cognitive information and its consequences for the project. Automation of tasks could eventually happen as we understand the process and the influence of contextual and cognitive utterances and differences. The goal is to automate the communication so
that developers do not have to rely on their memories, mental maps or searching huge sets of documentation. The chatbot would act upon a context model and would have embedded machine learning and history information to provide results to the developers. This integration would transform the chatbot into an intelligent tool. The solution is also intended to be non-invasive, relying on the implementation of techniques such as aggregated data or anonymization. With respect to the feasibility of implementation, our ongoing work on process-aware conversational agents has demonstrated that it is possible to integrate a basic context model into a chatbot tool such as RASA. We have also implemented a preliminary integration of the chatbot with a workflow machine called Camunda, allowing the chatbot to receive process execution information.

\vspace{0.2cm}
\noindent\fbox{
 \parbox{0.45\textwidth}{
 \textit{
Relevant questions include: How can we establish a communication channel between developers and their environment? How can a chatbot be effective in supporting developers? In which ways are developers willing to accept this new technology?
}
 }
}
\vspace{0.2cm}

\textbf{Knowledge-Intensive Process Guidance.} 
Knowledge workers such as software developers rely on their minds and creativity to implement software solutions. Providing smart solutions when building software is expected, so developers must worry about following patterns, processes and adjusting to project needs. This information is usually implicit or tacit and in developers’ memories. Automated task guidance or task navigation support should improve developers’ ability to work more efficiently. This system would count on intelligence from machine learning and project history to recognize the context in which developers are working and suggest the next steps, according to their context. Development tasks are knowledge guided, capturing the context is essential. Providing feedback to developers with recommendations should be part of the solution. Once the context is captured and understood, developers have a way to take advantage of this context through the chatbot. A method for process navigation recommendations based on machine-learning should be provided. The method may include training a machine-learning model by at least processing training data (context and process) with the machine-learning model. The training data may include records of the executed process with the current context at the time of execution. Correlations between context and the executed process shall be done so inferences of the next steps can be provided (recommended) to developers.

\vspace{0.2cm}
\noindent\fbox{
 \parbox{0.45\textwidth}{
 \textit{ Relevant questions include: How to leverage process information to guide developers in what they must do? What types of recommendations related to the software process can improve SD? How can developers take advantage of guidance recommendations? How can chatbot inferences be integrated with ML inferences to produce rich feedback to developers? How could we maximize the guidance of developers through process and context exploration?}
 }
}
\vspace{0.2cm}

\textbf{Experimental studies.} Qualitative and quantitative studies 
to demonstrate the feasibility of the proposed approaches as well as the implementation in software companies. Application areas of interest for the studies include deployment, testing, version control and managing issues or tasks.

\vspace{0.2cm}
\noindent\fbox{
 \parbox{0.45\textwidth}{
 \textit{
Relevant questions include: How can we verify the acceptance of a new cognitive assistance-based way of development by software developers? In which types of interactions would developers be mostly interested? How can we have confidence this idea will do what developers want?
}
 }
}
\vspace{0.2cm}

\section{Conclusion}

In this paper, we present and discuss a novel paradigm to improve the work of software developers, by providing contextual information about the tasks they are performing through cognitive and intelligent support. It employs capturing the implicit or tacit context and feeding it back in a useful way through recommendations to developers in real-time as they are executing the software project. The novelty of the paradigm arises from the approaches and tools used to capture and recommend the developers’ context on-the-fly, considering different contexts. The findings will be of interest since their use should have several advantages including less time to develop software, less effort to share knowledge among team members, enhanced collaboration, application of collective wisdom, knowledge transfer from experts to novice and many other useful contributions. As broader implications of the results, we believe that the impact of the proposed research will contribute to facilitating the development of new avenues in software research as well as support improved ways to develop software, a critical area that is in high demand and has enormous future growth potential. This is the first research program where the combination of three different pillars (context, chatbots and machine learning for process navigation) has been exploited to predict appropriate recommendations during software development.

\bibliographystyle{IEEEtran}
\bibliography{refs}

\begin{thebibliography}{10}
\providecommand{\url}[1]{#1}
\csname url@samestyle\endcsname
\providecommand{\newblock}{\relax}
\providecommand{\bibinfo}[2]{#2}
\providecommand{\BIBentrySTDinterwordspacing}{\spaceskip=0pt\relax}
\providecommand{\BIBentryALTinterwordstretchfactor}{4}
\providecommand{\BIBentryALTinterwordspacing}{\spaceskip=\fontdimen2\font plus
\BIBentryALTinterwordstretchfactor\fontdimen3\font minus
  \fontdimen4\font\relax}
\providecommand{\BIBforeignlanguage}[2]{{%
\expandafter\ifx\csname l@#1\endcsname\relax
\typeout{** WARNING: IEEEtran.bst: No hyphenation pattern has been}%
\typeout{** loaded for the language `#1'. Using the pattern for}%
\typeout{** the default language instead.}%
\else
\language=\csname l@#1\endcsname
\fi
#2}}
\providecommand{\BIBdecl}{\relax}
\BIBdecl

\bibitem{MEYER2017}
\BIBentryALTinterwordspacing
A.~N. Meyer, L.~E. Barton, G.~C. Murphy, T.~Zimmermann, and T.~Fritz,
  ``\BIBforeignlanguage{English}{The work life of developers: Activities,
  switches and perceived productivity},''
  \emph{\BIBforeignlanguage{English}{IEEE Transactions on Software
  Engineering}}, vol.~43, no.~12, pp. 1178--1193, 2017. [Online]. Available:
  \url{http://ieeexplore.ieee.org/document/7829407}
\BIBentrySTDinterwordspacing

\bibitem{Murphy_Beyond2019}
G.~Murphy, ``Beyond integrated development environments: adding context to
  software development,'' in \emph{Proceedings of the 41st International
  Conference on Software Engineering}.\hskip 1em plus 0.5em minus 0.4em\relax
  IEEE Press, 2019, pp. 73--76.

\bibitem{Ciccio2015}
C.~Di~Ciccio, A.~Marrella, and A.~Russo,
  ``\BIBforeignlanguage{English}{Knowledge-intensive processes:
  Characteristics, requirements and analysis of contemporary approaches},''
  \emph{\BIBforeignlanguage{English}{Journal on Data Semantics}}, vol.~4,
  no.~1, pp. 29--57, Mar 2015.

\bibitem{gronau2005kmdl}
N.~Gronau, C.~M{\"u}ller, and R.~Korf, ``Kmdl-capturing, analysing and
  improving knowledge-intensive business processes.'' \emph{J. UCS}, vol.~11,
  no.~4, pp. 452--472, 2005.

\bibitem{murphy_theneed2018}
\BIBentryALTinterwordspacing
G.~C. Murphy, ``The need for context in software engineering (ieee cs harlan
  mills award keynote),'' in \emph{Proceedings of the 33rd ACM/IEEE
  International Conference on Automated Software Engineering}, ser. ASE
  2018.\hskip 1em plus 0.5em minus 0.4em\relax New York, NY, USA: Association
  for Computing Machinery, 2018, p.~5. [Online]. Available:
  \url{https://doi.org/10.1145/3238147.3241987}
\BIBentrySTDinterwordspacing

\bibitem{Kersten_Murphy_2006}
\BIBentryALTinterwordspacing
M.~Kersten and G.~C. Murphy, ``Using task context to improve programmer
  productivity,'' in \emph{Proceedings of the 14th ACM SIGSOFT International
  Symposium on Foundations of Software Engineering}, ser. SIGSOFT
  ’06/FSE-14.\hskip 1em plus 0.5em minus 0.4em\relax ACM, 2006, p. 1–11.
  [Online]. Available: \url{http://doi.acm.org/10.1145/1181775.1181777}
\BIBentrySTDinterwordspacing

\bibitem{Gasparic_Murphy_Ricci_2017}
M.~Gasparic, G.~C. Murphy, and F.~Ricci, ``A context model for ide-based
  recommendation systems,'' \emph{Journal of Systems and Software}, vol. 128,
  p. 200–219, Jun 2017.

\bibitem{Holmes_Murphy_2005}
\BIBentryALTinterwordspacing
R.~Holmes and G.~C. Murphy, ``Using structural context to recommend source code
  examples,'' in \emph{Proceedings of the 27th International Conference on
  Software Engineering}, ser. ICSE ’05.\hskip 1em plus 0.5em minus
  0.4em\relax ACM, 2005, p. 117–125, event-place: St. Louis, MO, USA.
  [Online]. Available: \url{http://doi.acm.org/10.1145/1062455.1062491}
\BIBentrySTDinterwordspacing

\bibitem{devy}
N.~Bradley, T.~Fritz, and R.~Holmes, ``Context-aware conversational developer
  assistants,'' in \emph{2018 IEEE/ACM 40th International Conference on
  Software Engineering (ICSE)}.\hskip 1em plus 0.5em minus 0.4em\relax IEEE,
  2018, pp. 993--1003.

\bibitem{lima2015}
A.~M. {Lima}, R.~Q. {Reis}, and C.~A.~L. {Reis}, ``Empirical evidence of
  factors influencing project context in distributed software projects,'' in
  \emph{2015 IEEE/ACM 2nd International Workshop on Context for Software
  Development}, 2015, pp. 6--7.

\bibitem{biegel2015}
B.~{Biegel}, S.~{Baltes}, I.~{Scarpellini}, and S.~{Diehl}, ``Code basket:
  Making developers' mental model visible and explorable,'' in \emph{2015
  IEEE/ACM 2nd International Workshop on Context for Software Development},
  2015, pp. 20--24.

\bibitem{kidd1994marks}
A.~Kidd, ``The marks are on the knowledge worker,'' in \emph{Proceedings of the
  SIGCHI conference on Human factors in computing systems}, 1994, pp. 186--191.

\bibitem{Kersten2005}
M.~Kersten and G.~C. Murphy, ``Mylar: a degree-of-interest model for ides,'' in
  \emph{Proceedings of the 4th international conference on Aspect-oriented
  software development}.\hskip 1em plus 0.5em minus 0.4em\relax ACM, 2005, pp.
  159--168.

\bibitem{Cubranic2004Learning}
\BIBentryALTinterwordspacing
D.~\v{C}ubrani\'{C}, G.~C. Murphy, J.~Singer, and K.~S. Booth, ``Learning from
  project history: A case study for software development,'' in
  \emph{Proceedings of the 2004 ACM Conference on Computer Supported
  Cooperative Work}, ser. CSCW '04.\hskip 1em plus 0.5em minus 0.4em\relax New
  York, NY, USA: ACM, 2004, pp. 82--91. [Online]. Available:
  \url{http://doi.acm.org/10.1145/1031607.1031622}
\BIBentrySTDinterwordspacing

\bibitem{Cub_Murphy_2003}
\BIBentryALTinterwordspacing
D.~Čubranić and G.~C. Murphy, ``Hipikat: Recommending pertinent software
  development artifacts,'' in \emph{Proceedings of the 25th International
  Conference on Software Engineering}, ser. ICSE ’03.\hskip 1em plus 0.5em
  minus 0.4em\relax IEEE Computer Society, 2003, p. 408–418, event-place:
  Portland, Oregon. [Online]. Available:
  \url{http://dl.acm.org/citation.cfm?id=776816.776866}
\BIBentrySTDinterwordspacing

\bibitem{Ponzanelli_Bavota_Penta_Oliveto_Lanza_2014}
L.~Ponzanelli, G.~Bavota, M.~D. Penta, R.~Oliveto, and M.~Lanza, ``Prompter: A
  self-confident recommender system,'' in \emph{2014 IEEE International
  Conference on Software Maintenance and Evolution}, Sep 2014, p. 577–580.

\bibitem{cai2019answerbot}
L.~Cai, H.~Wang, B.~Xu, Q.~Huang, X.~Xia, D.~Lo, and Z.~Xing, ``Answerbot: an
  answer summary generation tool based on stack overflow,'' in
  \emph{Proceedings of the 2019 27th ACM Joint Meeting on European Software
  Engineering Conference and Symposium on the Foundations of Software
  Engineering}, 2019, pp. 1134--1138.

\bibitem{carzaniga1998characterization}
A.~Carzaniga, A.~Fuggetta, R.~S. Hall, D.~Heimbigner, A.~Van Der~Hoek, and
  A.~L. Wolf, ``A characterization framework for software deployment
  technologies,'' Colorado State Univ Fort Collins Dept of Computer Science,
  Tech. Rep., 1998.

\bibitem{Glaucia2019}
G.~Melo, T.~Oliveira, P.~Alencar, and D.~Cowan,
  ``\BIBforeignlanguage{English}{Retrieving curated stack overflow posts from
  project task similarities},'' in
  \emph{\BIBforeignlanguage{English}{Proceedings of the 31st International
  Conference on Software Engineering \& Knowledge Engineering}}, 2019, pp.
  415--418.

\bibitem{melo2019context}
G.~{Melo}, P.~{Alencar}, and D.~{Cowan}, ``Context-augmented software
  development in traditional and big data projects: Literature review and
  preliminary framework,'' in \emph{2019 IEEE International Conference on Big
  Data (Big Data)}, 2019, pp. 3449--3457.

\end{thebibliography}

\end{document}